\documentstyle[amssymb,multicol,prb,aps,epsfig]{revtex}
\begin{document}
\draft
\title{Low temperature transport through a quantum dot: \\
       finite-$U$ results and scaling behavior}
\author{D.\ Gerace, E.\ Pavarini, and L. C.\ Andreani}
\address{Istituto Nazionale per la Fisica della Materia
         and Dipartimento di Fisica ``A. Volta'',
         Universit\`a di Pavia, Via Bassi 6, 27100 Pavia, Italy}

\date{\today}
\maketitle
\begin{abstract}
We calculate the conductance through a quantum dot weakly coupled
to metallic leads, modeled by the spin-1/2 Anderson model with
finite Coulomb repulsion $U$. We adopt the non-crossing
approximation method in its finite-$U$ extension (UNCA). Our
results can be compared to those obtained with the exact numerical
renormalization group method, and good agreement is found both in
the high temperature (Coulomb blockade) and in the low temperature
(Kondo) regime. We analyze the scaling properties of the low
temperature conductance, and calculate the universal function
which describes the electronic transport in the Kondo regime. Very
good agreement with recent experimental results is found. Finally,
we suggest a simple interpolating function which fits fairly well
the calculated conductance in a broad temperature range.
\end{abstract}
\vskip 1cm

\begin{multicols}{2}
 
\section{Introduction}

The Kondo effect, a phenomenon discovered in the late 30s in
diluted magnetic alloys, plays a crucial role in the low
temperature properties of many strongly correlated systems, such
as Ce, Y and U compounds.\cite{hewson} Very recently Kondo-like
phenomena were also observed in the low-temperature transport
properties of quantum dot devices,
\cite{gold1,gold2,cronenwet,schmid,simmel,vanderwiel,schmid2,sasaki}
opening new opportunities to control the Kondo effect
experimentally, and starting a new field of research.\cite{kouw}

A quantum dot device consists of a small sized quantum dot (QD)
weakly connected by tunnel barriers to two electrodes, called
source and drain. In this system the QD may be considered as an
artificial atom, in which a well defined number of electrons, $n$,
is confined. The energy required to add a new electron to the QD
is $U=e^2/2C$, where $C$ is the capacitance of the QD itself. The
energy $U$ is determined by the Coulomb repulsion between two
electrons in the QD and thus it scales with the inverse of the dot
dimensions. Therefore, in small sized QD, $\; U$ is usually larger
than the coupling to the leads. An important aspect of a QD device
is that the energy of electrons in the dot can be tuned by a gate
voltage. Increasing the gate voltage, the energy necessary to add
a new electron decreases, eventually making the addition possible.
Thus it was shown that, at low temperature, the transport of
electrons through the QD is allowed only at those values of the
gate voltage at which the state with $n$ electrons is realized and
that with $n+1$ electrons becomes suddenly energetically
accessible (Coulomb blockade).\cite{been,meirav}

The idea that Kondo-like phenomena should appear in such a system
at very low temperature can be traced back to 1988. In that year
it was recognized that the Anderson model, introduced in 1961 to
describe a magnetic impurity in a metal,\cite{anderson} could also
be applied to a QD coupled to its leads.\cite{ng,glazman} Soon
afterwards several theoretical studies have been devoted to
analyze the properties of such a system (see e.g. Refs.
\onlinecite{nodd,meir91,meir92,meir93,meir94,izumida,izumida2,costi2}).
It was predicted that at very low temperature ($T\ll T_K$, where
T$_K$ is the Kondo temperature) a narrow peak should appear in the
local density of states, close to the Fermi level. Thus, states
belonging to opposite electrodes should mix easier than at high
temperature ($T\gg T_K$), and the conductance should increase. In
addition, the Kondo anomalies were predicted to appear only for
odd $n$, \cite{ng,glazman,nodd} and thus only if the total spin of
the electrons in the QD, $S$, is half integer. These predictions
are now confirmed by several experimental
results.\cite{gold1,gold2,cronenwet,schmid,simmel,vanderwiel} The
Kondo temperature of QD devices is however very small (usually
less than a few hundred mK) compared to the one of diluted
magnetic alloys (usually a few K). Recently, \cite{schmid2,sasaki}
in some QD devices Kondo anomalies have been observed for even $n$
as well; it is believed that these deviations from the even/odd
$n$ effect are related to the formation of integer spin states
with $S\ge1$.

In the present paper we will focus our attention on those QD
devices for which the Kondo effect occurs {\it only} for odd $n$.
The transport properties of such a QD device have been calculated
by using many different approaches, such as the equations of
motion method,\cite{meir91} the non-crossing approximation
(NCA)\cite{meir93,meir94} and the exact numerical renormalization
group (NRG) technique.\cite{izumida,izumida2,costi2} Within the
NRG it was possible to calculate the conductance as a function of
the gate voltage, and thus to obtain -- in the case $S=1/2$ -- the
even/odd effect in very good agreement with experimental results.
Similar results were also obtained with the approximate equations
of motion method and with the NCA. The latter
method\cite{cox,bickers} was applied to low temperature transport
of a QD in the infinite $U$ limit and thus the even/odd
alternation effect was obtained\cite{meir94} for $n$ switching
from $n=0$ to $n=1$.

Even if approximated, the NCA can be easily extended to models in
which the realistic levels structure of the QD is kept into
account (i.e. in the presence of several levels and possibly of an
external magnetic field), thus becoming a very convenient method
for a direct comparison to transport properties of real QD
devices. As a matter of fact, the NCA was used with success to
study, e.g., orbital degeneracy effects in diluted magnetic
alloys.\cite{cox,bickers,monnier} However, in the case of QD
devices,  it is first necessary to go beyond the
$\,U\longrightarrow\infty$ condition, which does not account for
the even/odd effect for all values of $n$.

In the present work we apply the finite-$U$ NCA
method\cite{pruschke} to the QD device system, modeled by the
spin-1/2 Anderson model. We calculate the conductance as a
function of the position of the QD level, and study the system in
the empty orbital, mixed valence and Kondo regimes. Our results
are in good agreement with the exact numerical renormalization
group ones. In particular, we obtain the correct behavior of the
conductance both in the high temperature regime, $T\gg T_K$, in
which the conductance shows the Coulomb blockade peaks, and the
low temperature regime, $T<T_K$, where the conductance increases
for odd $n$ due to the Kondo effect.

The paper is organized as follows. In section II we describe the
model Hamiltonian and give a short introduction to the finite-$U$
NCA. In section III we show and discuss the numerical results, in
particular the linear-response conductance, which we compare to
exact results and to experimental data. In section IV we study the
scaling behavior of the conductance as a function of temperature,
and we suggest an empirical formula which is valid both in the
Fermi-liquid and in the $T\sim T_K$ temperature regime, where the
conductance has a logarithmic behavior.

\section{Model and method}

\subsection{Hamiltonian}
In order to describe a quantum dot (QD) coupled with its leads we
adopt the Anderson Hamiltonian\cite{glazman}

\begin{eqnarray}
\label{H:anderson}
 H&=&\sum_{({\mathbf k},\sigma)\in S,D} \varepsilon_{\mathbf k}\,
        c_{{\mathbf k}\sigma}^{\dagger}
        c_{{\mathbf k}\sigma}+
\varepsilon_0\,\sum_{\sigma}\,d_{\sigma}^{\dagger}d_{\sigma} +U\,
n_{d\uparrow} n_{d\downarrow}
\nonumber \\
&+&\,\sum_{({\mathbf k},\sigma)\in S,D }
      \left(V_{{\mathbf k} \sigma}
            c_{{\mathbf k}\sigma}^{\dagger}
            d_{\sigma}+ h.c.
      \right).
\end{eqnarray}

Here $c_{{\mathbf k}\sigma}^\dagger$ ($c_{{\mathbf k}\sigma}$)
creates (destroys) a conduction electron with momentum {\bf k} and
spin $\sigma$ in one of the two leads, which we label with S
(source) and D (drain); $d_{\sigma}^\dagger$ ($d_{\sigma}$)
creates (destroys) an electron with
 spin $\sigma$ on the quantum dot;
$\varepsilon_0$ is the energy of a single electron localized on
the QD and $U$ is the Coulomb interaction among electrons in the
dot; $n_{d\sigma}=d_{\sigma}^{\dagger}d_{\sigma}$ is the number of
electrons operator for a given spin in the QD; { 
$V_{{\mathbf k} \sigma}$ is the
hybridization between the leads and the QD states,  whose
modulus is supposed to be {\bf k}-independent, 
with $|V_{{\mathbf k} \sigma}|=V_{S (D)}$
for ${({\mathbf k},\sigma)\in S(D)}$.
The lead-dot coupling strengths are given by  

\begin{equation}
\Gamma_{S(D)} \equiv \pi 
V_{S(D)}^2
\sum_{{\mathbf k}} 
\delta(\varepsilon-\varepsilon_{\mathbf k}).
\end{equation}

\noindent 
It was shown\cite{glazman} 
-- through a unitary transformation of the band states -- 
that the Hamiltonian (\ref{H:anderson})
is equivalent to a two band Anderson model in which
the first band does not interact with the QD,
and the second is coupled to the quantum dot states 
through the hybridization $V=\sqrt{V_S^2+V_D^2}$.
Thus the problem of calculating the 
transport properties of Hamiltonian (\ref{H:anderson})
is reduced to the problem of calculating the 
spectral properties of the one band Anderson model,
provided that the actual lead-dot coupling strength
is given by\cite{nota:gamma}
}

\begin{equation}
\label{gammatot} \Gamma=\Gamma_{S}+\Gamma_{D}=\pi
N(\varepsilon_F)\,V^2,
\end{equation}
where $N(\varepsilon_F)$ is the density of states (per spin) of
the leads at the Fermi level, $\varepsilon_F$. It is reasonable to
approximate the conduction bands of the leads with those of a non
interacting two dimensional Fermi gas. Thus the density of states
(DOS) per spin may be written as $N(\varepsilon)=1/2D$, where $D$
is one half of the band-width, therefore $\Gamma=\pi V^2/2D$.

The main difference between the Anderson model used in the
ordinary Kondo problem and the one used here is the following. In
the present model the energy difference
$\varepsilon_0-\varepsilon_F$ is not fixed, but, on the contrary,
it can be tuned by a gate voltage $V_g$, coupled to the QD through
a capacitor. From now on we set the Fermi level $\varepsilon_F=0$.
In first approximation, we can assume that $-\varepsilon_0$
increases linearly with $eV_g$ (with common conventions for the
sign of gate voltage). Thus the equilibrium thermodynamic
properties of the system described by the Hamiltonian
(\ref{H:anderson}) are functions of two external parameters: the
temperature $T$ and the gate voltage $V_g$.

\subsection{Linear response conductance}

In the linear response regime 
($V_{SD}\ll V_g$, where $V_{SD}$ is the source-drain voltage), the conductance
of the system QD+leads, G, may be written in a Landauer-like
form\cite{meir91,meir92}

\begin{equation}
\label{landauer} {\rm
G}(T,V_g)\!=\!\frac{2e^2}{h}\!\int_{-\infty}^{+\infty}
\!\!\!\!\!\!\!\!\!\pi\Gamma\left(-
\frac{1}{\pi}{\rm Im}\{G^R(\varepsilon+{\rm i}\eta)\}\right)
\left(-\frac{\partial{f}}{\partial{\varepsilon}}\right)\,{\rm
d}\varepsilon,
\end{equation}
where for simplicity we assume that the couplings to the leads are
symmetric ($\Gamma_S=\Gamma_D$). Here $f$ is the Fermi
distribution function and $G^R(\varepsilon+{\rm i}\eta)$ 
is the retarded local Green function, 
{ i.e. the Fourier transform
of  $G^R(t)\equiv - \mbox{\rm i} \Theta(t) \langle \{ d(t),d^\dag (0)\} 
\rangle$ }. 

The quantity Im $\left\{
G^R(\varepsilon+{\rm i}\eta) \right\} $ in Eq. (\ref{landauer}) is
proportional to the local density of states, $\rho(\varepsilon)$,
defined as

\begin{equation} \label{dosanders}
\rho(\varepsilon)\equiv-\frac{1}{\pi}\,{\rm Im}\left\{G^{R}
(\varepsilon +{\rm i}\eta)\right\}.
\end{equation}
We point  out that the effects of the Coulomb interaction among
electrons in the QD are contained in the local density of states.
The main purpose of this work is thus to calculate
$\rho(\varepsilon)$ as a function of temperature and gate voltage,
and the resulting linear conductance.

\subsection{Number of electrons}

The low temperature transport properties of a mesoscopic system
such as the QD+leads previously described are characterized by the
phenomenon of charge quantization. When Coulomb blockade occurs
the number of electrons in the QD is a fixed integer. This number
can be changed by raising the voltage of the gate electrode. In
this way the energy of the electrons in the QD is lowered with
respect to the Fermi energy in the leads. The change in energy
necessary to add a new electron is $\sim U$. This may be seen as
the energy required to charge the QD, $e^2/2C$, where $C$ is the
capacitance of the QD. Thus the average number of electrons on the
QD, $\langle n \rangle$, with

\begin{equation}\label{sumrule}
\langle n\rangle  =N\;\int_{-\infty}^{+\infty}{\rm d}
\varepsilon\, f(\varepsilon)\rho(\varepsilon),
\end{equation}
is an important quantity in this problem (here $N$ is the
degeneracy of $\varepsilon_0$ and in the present case $N=2$). The
number of electrons is --- like the conductance --- a function of
the local density of states.

\subsection{Finite-$U$ non-crossing approximation}

In order to calculate $\rho(\varepsilon)$, the local DOS for the
single impurity Anderson model, we adopt the finite-$U$
non-crossing approximation (UNCA) approach,\cite{pruschke} an
extension of the non-crossing approximation (NCA) to the finite
$U$ Anderson model. We point out that a finite $U$ treatment is
necessary in order to calculate the correct behavior of the
conductance as a function of the gate voltage (and thus of the
number of electrons on the QD).

The NCA is a diagrammatic technique introduced for the Anderson
model in the $U\!\longrightarrow\!\infty$ limit.\cite{cox,bickers}
The main idea of the NCA is that the self energy can be expanded
in a series of diagrams of order $V^2/N$, where $N$ is the
degeneracy of the local level, usually large for Kondo impurities
in diluted magnetic alloys (e.g. $N=6$ for Ce impurities). The
non-crossing diagrams are summed up to all orders, and the first
neglected diagrams are of order $(V^2/N)^2$. The NCA was proved to
be successful\cite{cox2} already for $N=2$. Thus it was applied
with good results also to the spin-1/2 $\,U\rightarrow\infty$
Anderson model out of equilibrium,\cite{meir93,meir94} e.g. it was
used to calculate the behavior of the conductance as a function of
the chemical potential in the QD+leads problem.

The NCA and the UNCA lead to a set of integral equations for the
self energy, which have to be solved self-consistently. The number
of equations depends on the number of many body states for the
electrons in the QD. If $U$ is finite there are four possible
manybody states, namely $|0,0\rangle$  (with energy
$\varepsilon=0$), $|1,\downarrow\rangle$ and $|1,\uparrow\rangle$
(degenerate, with energy $\varepsilon=\varepsilon_0$), and
$|2,\uparrow\downarrow \rangle$ (with energy $\varepsilon=
2\varepsilon_0+U$). For each state a self energy and a ionic
propagator are introduced. Thus  we have  three self energies,
$\Sigma_0$, $\Sigma_1$, $\Sigma_2$ and three propagators, $G_0$,
$G_1$, $G_2$ (due to spin degeneracy
$G_{1\uparrow}=G_{1\downarrow}=G_1$ and
$\Sigma_{1\uparrow}=\Sigma_{1\downarrow}=\Sigma_1$). The UNCA
equations may then be written as\cite{pruschke,schmalian}

\begin{eqnarray}
\Sigma_0(\omega+{\rm
i}\eta)&=&\frac{2\Gamma}{\pi}\int_{-D}^{D}{\rm d}\varepsilon \;
f(\varepsilon)G_1(\omega+\varepsilon+{\rm i}\eta)
,\label{eq1}\\
\Sigma_1(\omega+{\rm i}\eta)&=&\frac{\Gamma}{\pi}\int_{-D}^{D}{\rm
d}\varepsilon \; [1-f(\varepsilon)]G_0(\omega-\varepsilon+{\rm i}\eta)+\nonumber\\
& & +\frac{\Gamma}{\pi}\int_{-D}^{D}{\rm d}\varepsilon \;
f(\varepsilon)G_2(\omega+\varepsilon+{\rm i}\eta)
,\label{eq2}\\
\Sigma_2(\omega+{\rm
i}\eta)&=&\frac{2\Gamma}{\pi}\int_{-D}^{D}{\rm d}\varepsilon \;
[1-f(\varepsilon)]G_1(\omega-\varepsilon+{\rm i}\eta) ,\label{eq3}
\end{eqnarray}
with

\begin{eqnarray}
G_0(\omega+{\rm i}\eta)&=&\left[\omega+{\rm
i}\eta-\Sigma_0(\omega+{\rm
i}\eta)\right]^{-1},\label{eq4}\\
G_1(\omega+{\rm i}\eta)&=&\left[\omega+{\rm i}\eta-\varepsilon_0-
\Sigma_1(\omega+{\rm
i}\eta)\right]^{-1},\label{eq5}\\
G_2(\omega+{\rm i}\eta)&=&\left[\omega+{\rm
i}\eta-2\varepsilon_0-U-\Sigma_2(\omega+{\rm
i}\eta)\right]^{-1}.\label{eq6}
\end{eqnarray}
In the infinite $U$ limit, the double occupation state is
forbidden, and $G_2$ may be neglected.

Once this system of six equations is solved, it is possible to
express all the physical quantities in terms of the ionic
resolvents. In particular the retarded local Green function may be
evaluated by analytic continuation from the corresponding
imaginary time propagator, which may be written as\cite{pruschke}

\begin{eqnarray}
\nonumber
G({\rm i}\omega)&=& \frac{1}{Z} \oint_C \frac{{\rm d}z} {2\pi{\rm
i}} \; \mbox{\rm e}^{-z/k_BT} \;\times\\
&&\left[ G_0(z) G_1(z+{\rm
i}\omega)+G_1(z)G_2(z+{\rm i}\omega)\right],
\end{eqnarray}
where $Z$ is the QD partition function, i.e.

\begin{equation}
Z= \oint_C \frac{{\rm d}z}{2\pi {\rm i}}\; \mbox{\rm e}^{-z/k_BT}
\; \left[G_0(z)+G_1(z)+G_2(z)\right]
\end{equation}
Finally, the local density of states of the QD can be obtained
from the retarded local Green function through Eq.
(\ref{dosanders}).

\begin{figure}[!ht]
\begin{center}
\includegraphics[width=0.48\textwidth]{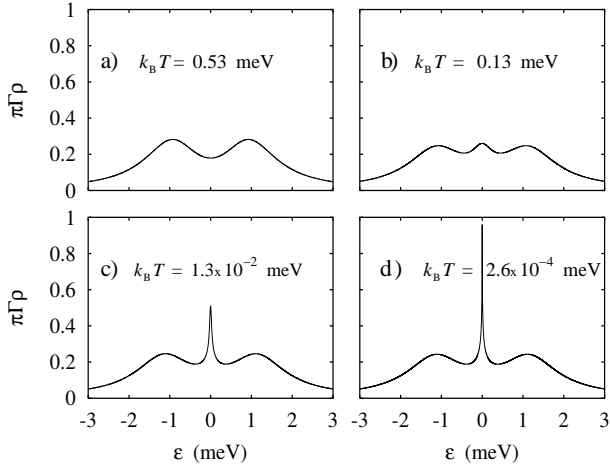}
\\[1ex]
\caption{ Equilibrium density of states for the particle-hole
symmetric Anderson model ($U=-2\varepsilon_0$) at different
temperatures. Parameters: $\varepsilon_0=-1$ meV, $U=2$ meV,
$\Gamma=0.35$ meV, $D=4$ meV. For $k_BT=2.6\times 10^{-4}$ meV
($T=3$ mK) the local DOS at Fermi energy is
$\pi\Gamma\rho(0)=0.98$. In this paper the Fermi level
$\varepsilon_F$ is set equal to zero.}\label{doskondo}
\end{center}
\end{figure}

The NCA and the UNCA break down below a temperature $T^*$, which
in the Kondo regime is much smaller than
$T_K$.\cite{cox,bickers,monnier,pruschke} Below this temperature
spurious features show up, e.g., in the local density of states.
It is known that the exact results are recovered with the
inclusion of vertex corrections,\cite{anders,costi1,haule} within
a set of integral equations which are numerically heavier than
Eqs. (\ref{eq1}-\ref{eq6}). This is, however, beyond the purpose
of the present paper. Here we will use the UNCA without vertex
diagrams. We will see that, nevertheless, we can reproduce fairly
well both the high and low temperature regimes, and that (above
$T^*$) the local DOS calculated with the UNCA is in good agreement
with the exact numerical renormalization group (NRG) results.

\section{Numerical results} \label{sec3}

We solve the self-consistent UNCA equations, and calculate the
local DOS, by using the fast Fourier transforms
technique.\cite{schmalian} From $\rho(\varepsilon)$ we then obtain
the Landauer conductance and the number of electrons as a function
of the temperature and of the position of the QD level,
$\varepsilon_0$, proportional to $-V_g$.

\subsection{Local density of states}

The local density of states is shown in Fig. \ref{doskondo} as a
function of the energy. The ratio $-\varepsilon_0/\Gamma\sim
2.85>1$, and thus the model describes the system in the Kondo
regime. For simplicity, we show the results for the particle-hole
symmetric Anderson model ($U=-2\varepsilon_0$). The Kondo
temperature may be estimated from Haldane formula,\cite{haldane}

\begin{equation}
\label{haldane} k_BT_K\sim (U\Gamma/2)^{1/2} \; \mbox{\rm exp}
\left[ \pi \varepsilon_0 (\varepsilon_0+U)/ (2 U\Gamma) \right]
,\end{equation} and in the present case we find $k_BT_K\sim 0.06$
meV.

In the high temperature limit ($T\gg T_K$, Fig. \ref{doskondo}a),
the local density of states shows two broad resonances. The first
one is located at energy $\varepsilon\!=\!
E(n=1)\!-\!E(n=0)\!\equiv\!\varepsilon_0$, the energy required to
put the first electron in the QD. The second resonance appears at
$\varepsilon\!=\!E(n=2)\!-\!E(n=1)\!\equiv\!\varepsilon_0+U$, the
energy required to add an additional electron. The shape of the
resonance peaks is Lorentzian. The origin of the peak broadening
is mainly the coupling QD-leads measured by the width $\Gamma$,
although a small thermal contribution is present since
$k_BT\!\gtrsim\! \Gamma$.

As the temperature decreases, a peak -- which is the fingerprint
of the Kondo effect -- appears close to the Fermi level. The
height of the peak increases on decreasing the temperature. Figure
\ref{doskondo} shows that for $T\ll T_K$ the height of the Kondo
peak tends to a maximum value, i.e.
$\pi\Gamma\rho(\varepsilon_F=0)\longrightarrow1$, which is reached
exactly only at $T=0$. At $k_BT=2.6\times 10^{-4}$ meV ($T/T_K\sim
0.01$, Fig. \ref{doskondo}d) we find $\pi\Gamma\rho(0)\sim 0.98$.

The evolution of $\rho(\varepsilon)$ as a function of temperature
is in fairly good agreement with the NRG
results.\cite{izumida,costi2,costi3} In the very large $U$ limit
we find results consistent with those obtained by using the NCA
method with infinite $U\;$.\cite{cox,bickers}

\subsection{Linear-response conductance}

\begin{figure}[!ht]
\begin{center}
\includegraphics[width=0.45\textwidth]{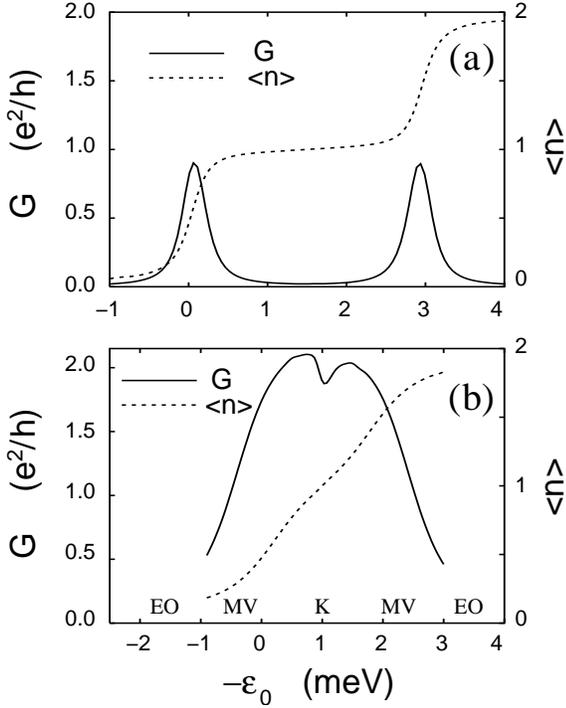}
\\[1ex]
\caption{Linear-response conductance G and average number of
electrons on the dot $\langle n \rangle$, for two different set of
parameters: (a)    Parameters are: $\Gamma=0.094$ meV, $D=4$ meV,
    $U=3$ meV,  and $k_BT=4.3\times 10^{-2}$ meV ($T=500$ mK)
       (Coulomb blockade);
(b)    Parameters are: $\Gamma=0.35$ meV,  $D=4$ meV,
    $U=2$ meV, and $k_BT=4.3\times 10^{-4}$ meV ($T=5$ mK).
       The various regimes as a function of $-\varepsilon_0$ are
       schematically indicated in (b) (see text).}\label{gelnum}
\end{center}
\end{figure}

The linear response conductance G and the average number of
electrons $\langle n \rangle$ are shown in Fig. \ref{gelnum} as a
function of $-\varepsilon_0$.\cite{nota} For $-\varepsilon_0<U/2$,
when the number of electrons in the dot is $n\leq 1$, there are
three relevant regimes of interest: the Kondo (K) regime
$\Gamma<-\varepsilon_0<U/2$, the mixed valence (MV) regime
$|\varepsilon_0|<\Gamma$, and the empty orbital (EO) regime
$-\varepsilon_0<-\Gamma$. In the opposite case
$-\varepsilon_0>U/2$, when the number of electrons in the QD is
$n\geq 1$, it is convenient to discuss the various regimes in
terms of holes; thus the Kondo regime exists for
$U/2<-\varepsilon_0<U-\Gamma$, the mixed valence regime for
$|-\varepsilon_0-U|<\Gamma$, and the empty orbital regime with
$n\simeq 2$ occurs for $-\varepsilon_0>U+\Gamma$. These regimes
are schematically indicated in Fig. \ref{gelnum}b.

In the Kondo regime, we can extimate the Kondo temperature from
Haldane formula, Eq. (\ref{haldane}). We find $k_BT_K\sim
10^{-5}-10^{-2}$ meV for $\Gamma\sim0.094$ (Fig. \ref{gelnum}a)
and $k_BT_K\sim 0.062-0.161$ meV for $\Gamma\sim 0.35$ (Fig.
\ref{gelnum}b).

Figure \ref{gelnum}a shows the results obtained for $k_BT\sim
0.5\Gamma\gg k_BT_K$. In this limit it was shown\cite{been,meir91}
that the conductance has narrow peaks every time the average
number of electrons in the QD increases by one. The peaks are
separated by valleys in which the conductance is almost zero. This
behavior (which is due to the Coulomb blockade\cite{meirav}) is
well reproduced in Fig. \ref{gelnum}a. The peaks reach about
\noindent
$e^2/h$, as expected from Coulomb blockade theory in the limit
$k_BT\sim\Gamma$, and have a width of about $2\Gamma$. This
broadening is due to tunneling, while the thermal broadening is
negligible. The line shape is almost Lorentzian. If the
temperature is raised so much that $k_BT\gg\Gamma$ the height of
the peaks becomes much smaller than $e^2/h$ and their width is
controlled by thermal broadening only.

Figure \ref{gelnum}b shows G and $\langle n \rangle$ for $T\ll
T_K$. In this regime, experiments
\cite{gold1,gold2,cronenwet,schmid,simmel,vanderwiel} show that
the valleys tend to raise when the number of electrons in the QD
is odd, and remain almost unchanged when the number of electrons
in the QD is even. In the very low temperature limit, the valleys
with an odd number of electrons evolve into a
plateau\cite{vanderwiel} at ${\rm G}\!\sim\! 2e^2/h$. This behavior is
the manifestation of the Kondo effect in the mesoscopic transport
properties of the QD+leads system. Figure \ref{gelnum}b shows that
when $\langle n \rangle\sim 1$ the conductance calculated with the
UNCA tends to ${\rm G}=2e^2/h$, while a valley appears for
$\langle n \rangle\sim 0$ and $\langle n \rangle\sim 2$. Thus the
Kondo regime is well described by the adopted method. The average
number of electrons in Fig. \ref{gelnum}b has a regular increase
and does not have well defined plateaux as in Fig. \ref{gelnum}a,
due to the larger value of~$\Gamma$.

Thus we have shown that the UNCA can describe well 
\begin{figure}[!ht]
\begin{center}
\includegraphics[width=0.45\textwidth]{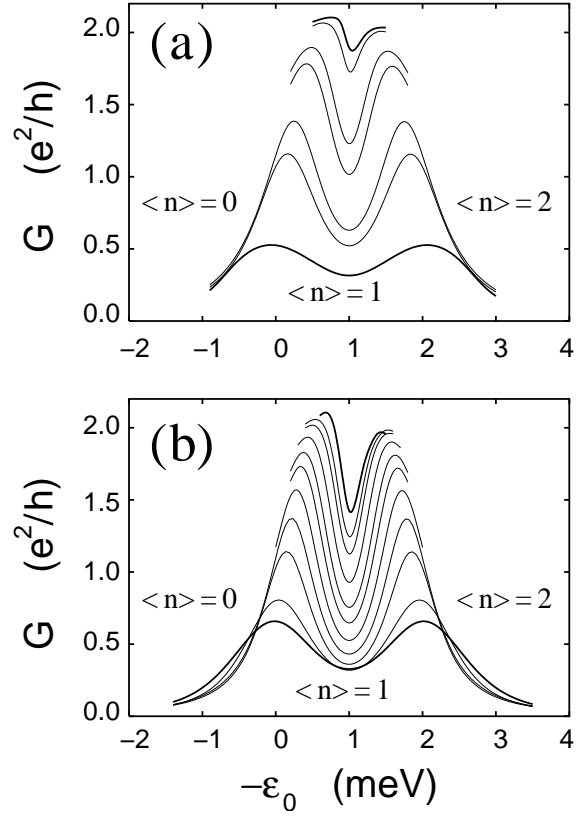}
\\[1ex]
\caption{Linear response conductance as a function of
$-\varepsilon_0$ ($\varepsilon_0$ is always referred to the Fermi
level, being $\varepsilon_F=0$) the energy of the dot level with
respect to the Fermi energy, and for many temperatures. (a)
Parameters: $\Gamma=0.35$ meV, $D=4$ meV, $U=2$ meV.
    Temperatures (from the bottom to the top curve):
    $k_BT=0.43$, 0.086, 0.043, $8.6\times 10^{-3}$,
          $4.3\times 10^{-3}$, $8.6\times 10^{-4}$,
           and $4.3\times 10^{-4}$ meV.
(b) Parameters: $\Gamma=0.275$ meV, $D=4$ meV, $U=2$ meV.
    Temperatures (from the bottom to the top curve):
    $k_BT=0.26$, 0.17, 0.07, 0.035, 0.017, $8.6\times 10^{-3}$,
          $5.3\times 10^{-3}$, $2.6\times 10^{-3}$,
          $1.3\times 10^{-3}$, $8.6\times 10^{-4}$,
          and $4.3\times 10^{-4}$ meV.}\label{ge}
\end{center}
\end{figure}
\noindent
both the
Coulomb blockade and the Kondo effect. The results can be compared
with the exact NRG calculations.\cite{izumida,izumida2,costi2}
Agreement is very good in the Kondo regime, down to $T\sim
0.01T_K$. In the mixed valence and empty orbital regimes the
agreement decreases. At low temperature the UNCA breaks down and a
spurious peak appears in the local DOS, close to the Fermi level,
as discussed for the usual NCA in Ref. \onlinecite{costi1}. Within
the parameters and the temperature chosen in Fig. \ref{gelnum}b,
this happens approximately for $-\varepsilon_0<0.5$ meV and
$-\varepsilon_0>1.5$ meV. In this region the conductance
calculated with the UNCA is much larger than the exact NRG result.
Therefore from now on we will plot the conductance only for the
values of $T$ and $\varepsilon_0$ for which the UNCA is reliable.

The linear-response conductance is shown in Fig. \ref{ge} as a
function of  $-\varepsilon_0$  and for several different
temperatures. The results are shown for two different choices of
the tunneling width: $\Gamma=0.35$ meV (Fig. \ref{ge}a) and
$\Gamma=0.275$ meV (Fig. \ref{ge}b). In the Kondo regime, we
estimate the Kondo temperature by using Haldane formula, and find
$k_BT_K\sim 0.062-0.161$ meV for $\Gamma=0.35$ meV and $k_BT_K\sim
0.030-0.134$ meV for $\Gamma=0.275$ meV.

Figure \ref{ge}a shows that at  high temperature ($k_BT>\Gamma\gg
k_BT_K$) there are two Coulomb blockade peaks at energy
$-\varepsilon_0\sim0$ and $-\varepsilon_0\sim U$, the maximum
value of the conductance being ${\rm G}\sim 0.5 e^2/h$. As the
temperature decreases, the Coulomb blockade peaks become closer to
each other, and ideally merge into a plateau at $T=0$. This is due
to the fact that the energy $-\varepsilon_0$ itself is
renormalized by the many body effects, through the real part of
the self energy. At very low temperature we see that ${\rm G}\sim
2e^2/h$ (with maximum deviation of $10\%$) in the region in which
the number of electrons is odd $(n\sim1)$.

The same behavior is shown in Fig. \ref{ge}b. Here we choose a
smaller value of $\Gamma$, so that the Kondo temperature is
smaller. At high temperature we see again the two Coulomb blockade
peaks, located at $-\varepsilon_0\sim 0$ and $-\varepsilon_0\sim
U$. At very low temperature the peaks approach each other and tend
to the maximum value $2e^2/h$. However in this case the deviation
from the limiting value is larger (about $25\%$). This is due to
the fact that, because of the smaller $T_K$, down at $T\sim 0.01
T_K$ (about the lowest temperature that can be reached before the
UNCA breaks down) we have $\pi\Gamma\rho(0)\sim 0.75$ instead of
$\pi\Gamma\rho(0)=1$. Nevertheless, Fig. \ref{ge}b reproduces the
theoretical NRG calculations in the Kondo regime at several
different temperatures, see e.g. Fig. 3 in Ref.
\onlinecite{izumida}, Fig. 2 in Ref. \onlinecite{izumida2}, and
Fig. 2a of Ref. \onlinecite{costi2}. In addition, at high
temperature Fig. \ref{ge}b provides very good results  in the
whole range of $-\varepsilon_0$.

It is worth to remind that when $k_BT>\Gamma$ the peak broadening
is mainly due to thermal effects. Thus, when we start from the
high temperature regime and lower the temperature, at first the
peak width decreases. This effect is shown in Fig. \ref{ge}b. On
the contrary, when $k_BT\sim\Gamma$ the broadening is about
$2\Gamma$, in agreement with experimental
results.\cite{gold2,vanderwiel} If the Kondo temperature is
negligibly small, lowering the temperature below $\Gamma/k_B$ does
not produce any further change in the peak width. However, if the
Kondo temperature is finite, as soon as $k_BT\sim k_BT_K\ll\Gamma$
the two peaks tend to become broader, and merge in a single large
peak at zero temperature. This effect is shown in Fig. \ref{ge}.

The UNCA results in Fig. \ref{ge} can be directly compared to the
experimental measurements of the linear conductance vs the gate
voltage. The curves in Fig. \ref{ge} reproduce very well the
experiments described in Refs.
\onlinecite{gold1,gold2,cronenwet,vanderwiel}, in which the
measurement of the Kondo effect in the equilibrium conductance of
a QD+leads system is reported. Agreement is satisfactory for the
temperature dependence and the broadening of the peaks, and it
covers the whole range of electron occupancy $0\leq n\leq 2$.

\begin{figure}[!ht]
\begin{center}
\includegraphics[width=0.4\textwidth]{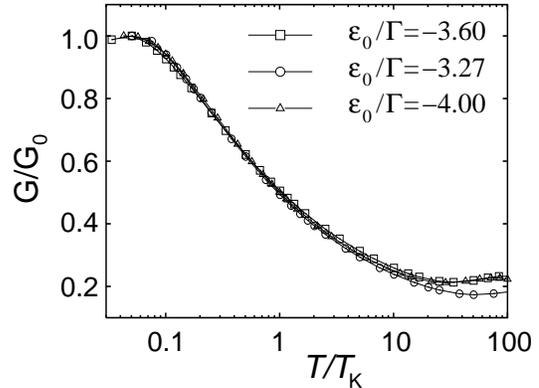}
\\[1ex]
\caption{ Universal behavior of linear-response conductance
normalized to its saturation value ${\rm G}_0$. The conductance is
displayed as a function of $T/T_K$, for three different values of
$\varepsilon_0/\Gamma$ chosen in the Kondo regime. Parameters:
$\Gamma=0.275$ meV, $D=4$ meV, $U=2$ meV.}\label{gtuniv}
\end{center}
\end{figure}

\section{Scaling behavior} \label{sec4}

In diluted magnetic alloys, the low temperature ($T\le T_K$)
resistivity, $\rho(T)$, follows a universal scaling
law,\cite{hewson,costi3} i.e. $\rho/\rho_0=F(T/T_K)$, where
$\rho_0\equiv \rho(T=0)$ and $F(x)$ is a function independent on
system-related parameters. This happens because at low temperature
and in the Kondo regime there is only one relevant energy scale,
$k_BT_K$. Similar scaling properties have been reported for the
low temperature conductance of QD
devices,\cite{gold2,vanderwiel,sasaki,kouw} and carbon
nanotubes.\cite{nygard} For a QD device described by the spin-1/2
Anderson model the universal function ${\rm G}/{\rm G}_0$ was
recently calculated by using the NRG.  \cite{izumida2} Here we
calculate the same function by using the UNCA, and compare our
results with the NRG calculations and experimental data.

\subsection{Universal curves}
The conductance calculated with the UNCA is shown in in Fig.
\ref{gtuniv} as a function of $T/T_K$. The different set of points
correspond to different choices of $-\varepsilon_0/\Gamma$ in the
Kondo regime ($-\varepsilon_0/\Gamma=3.6$ corresponds to the
particle-hole symmetric point, that is $-\varepsilon_0=U/2=1$ meV
in Fig. \ref{ge}b). The conductance is normalized to
G$_0\equiv$G$(T_0,V_g)$, where $T_0$ is the lowest temperature
which can be reached before the UNCA breaks down. In order to
achieve a universal behavior we define the Kondo temperature such
that G$(T_K)={\rm G}_0/2$, as often done.\cite{gold2,vanderwiel}
The relative deviation from ideal result in the Kondo region for
$-\varepsilon_0/\Gamma$, that is $\delta=(\mbox{\rm
G}_0-2e^2/h)/(2e^2/h)$, depends on the choice of the parameters
(we find $2\%<|\delta|<25\%$). Figure \ref{gtuniv} shows that the
conductance follows the universal behavior in a fairly large range
of temperatures. Departures from this universality are observed
only for $T\gg T_K$, as 
\begin{figure}[!ht]
\begin{center}
\includegraphics[width=0.45\textwidth]{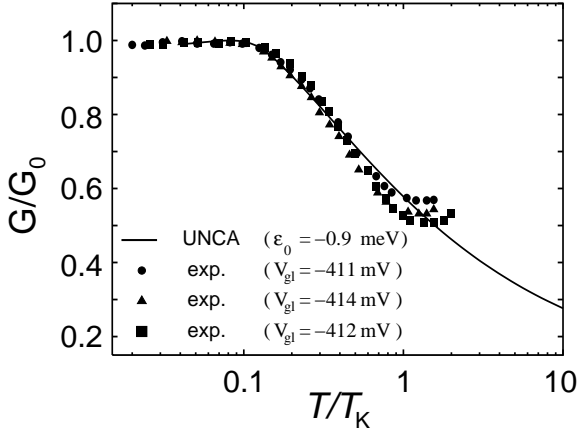}
\\[1ex]
\caption{Experimental data (Fig. 3c of Ref. 7)
         compared with the UNCA universal curve ${\rm G}/{\rm G}_0$.
         Parameters of the theoretical curve:
         $\varepsilon_0/\Gamma=-3.27$, $D=4$ meV, $U=2$ meV.
         Here $V_{gl}$ is the gate voltage.\cite{vanderwiel}}\label{fitvanderw}
\end{center}
\end{figure}
\noindent
expected. We point out that at very low
temperature ($T/T_K\ll 1$) the universal curve has the expected
Fermi liquid behavior,\cite{nazarov} i.e. $({\rm G}-{\rm
G}_0)/{\rm G}_0\propto T^2$, while at higher temperature
($T/T_K\sim 1$) the conductance is proportional to $\mbox{\rm ln }
(T/T_K)$.

The universal curve of the conductance calculated with the UNCA is
similar to NRG results for the ordinary Kondo problem (see e.g.
Fig. 4 of Ref. \onlinecite{gold2} or Fig. 12 of Ref.
\onlinecite{costi3}) and to NRG results for the conductance in QD
devices (see Figs. 4 and 7a of Ref.  \onlinecite{izumida2}).

\subsection{Comparison with experiments}

The universal curve calculated here can be directly compared with
existing measurements of the conductance. In this section we will
compare in particular with the experimental data taken from Ref.
\onlinecite{vanderwiel}. In Fig. \ref{fitvanderw} we show G$/{\rm
G}_0$ vs $T/T_K$, and plot both the experimental points and our
UNCA curve. In Ref. \onlinecite{vanderwiel} the experimental data
were normalized to ${\rm G}_0$, the value of the conductance at
the lowest temperature for which a measurement exists, and the
Kondo temperature was defined as the temperature such that
G$(T_K)={\rm G}_0/2$.

We found the best agreements between UNCA and experiments for the
choice $\varepsilon_0/\Gamma=-3.27$. Figure \ref{fitvanderw} shows
that the crossover from the logarithmic behavior to the low
temperature Fermi liquid regime is reproduced fairly well by the
conductance calculated by using the UNCA.

\subsection{Interpolation formula}

In order to extract the Kondo temperature from experimental data,
it has become very common\cite{gold2,vanderwiel,nygard} to fit the
data with the empirical formula
\begin{figure}[!ht]
\begin{center}
\includegraphics[width=0.45\textwidth]{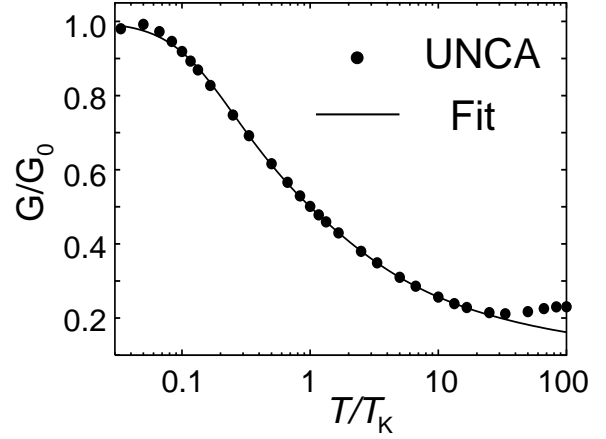}
\\[1ex]
\caption{Universal curve ${\rm G}/{\rm G}_0$
($\varepsilon_0/\Gamma=-3.6$) compared with the interpolation
function given in Eq. (\ref{fit3par}). The resulting fit
parameters are: $a=1/\ln(1+b+c)\simeq 0.25$, $b=43.4$ and
$c=10.25$. }\label{ncafit}
\end{center}
\end{figure}

\begin{equation}\label{fit1par}
{\rm G}(T)={\rm G}_0\left(\frac{T_K'\,^2}{T^2+T_K'\,^2}\right)^s
\end{equation}
where $T_K'=T_K/\sqrt{2^{1/s}-1}$ and G$(T_K)={\rm G}_0/2$. This
formula reproduces well NRG results for $T<T_K$ and has a single
fitting parameter, $s$. However, in the intermediate temperatures
regime ($T\sim T_K$) the exact conductance is proportional to
ln$(T/T_K)$, as NRG calculations show,\cite{izumida2} and this
behavior is not explicitly accounted for in formula
(\ref{fit1par}). Here we propose an alternative phenomenological
formula, which (a) is still quite simple, (b) reproduces fairly
well the calculated conductance in the whole temperature regime
and (c) shows explicitly the logarithmic behavior in the
intermediate temperature regime. Our formula has two fitting
parameters instead of one.

We notice that at very low temperature, in the Fermi liquid regime
($T\ll T_K$), the conductance should show the Fermi liquid $T^2$
behavior,

\begin{equation}\label{liquidofermi}
{\rm G}={\rm G}_0\left(1-\alpha\frac{T^2}{T_K^2}\right),
\end{equation}
where $\alpha$ is a parameter. Instead, at higher temperatures
($T\sim T_K$) a logarithmic behavior is expected

\begin{equation}\label{altatemp}
{\rm G}\propto {\rm G}_0\ln{(T/T_K)}.
\end{equation}
A function that satisfies both requirements is the following

\begin{equation}\label{fit3par}
\frac{{\rm G}}{{\rm
G}_0}=\left(1+a\ln{\left[1+b\left(\frac{T}{T_K}\right)^2+
c\left(\frac{T}{T_K}\right)^4\right]}\right)^{-1},
\end{equation}
where $a$, $b$, and $c$ are dimensionless parameters to be
determined with best fit techniques. By defining $T_K$ as the
temperature such that G$(T_K)={\rm G}_0/2$, we find
$a=1/\ln(1+b+c)$. Thus there are only two fitting parameters in
our formula.

In Fig. \ref{ncafit} we show the results of a fit for the choice
$-\varepsilon_0/\Gamma=-3.6$. The empirical formula reproduces
extremely well the UNCA results for the values $b=43.4$,
$c=10.25$. These parameters slightly depend, of course, on the
choice of $-\varepsilon_0/\Gamma$, and of the other UNCA
parameters. Thus Eq. (\ref{fit3par}) reproduces the correct
behavior, and it is physically correct both in the $T\ll T_K$ and
$T\sim T_K$ regime. We notice that the fourth order term is
required to reproduce the behavior of the conductance in the
intermediate temperature regime, while the $T^2$ term is required
to reproduce the low temperature Fermi liquid behavior.

\section{Conclusion} \label{sec5}

In the present work we have calculated the conductance of a system
made of a quantum dot coupled to two leads, described by the
spin-1/2 Anderson model. We adopted the finite-$U$ non-crossing
approximation method (UNCA), which allowed us to calculate the
conductance for the Anderson model with finite Coulomb repulsion.
Thus we were able to study the conductance as a function of
temperature and gate voltage. We have shown that the results
obtained with this method are in good agreement with those
obtained by using the exact numerical renormalization group
(NRG).\cite{izumida,izumida2,costi2} In addition we reproduced
both the Coulomb blockade and the Kondo effect in quantum dots.
Inclusion of a finite Coulomb correlation is important in order to
describe correctly the experimental results in the whole regime of
electron occupancies. The comparison with experimental data of
Ref. \onlinecite{vanderwiel} is fairly good, for what concerns
both temperature and gate voltage dependence. Finally we have
suggested a simple phenomenological formula which fits UNCA
results both in the logarithmic and in  the Fermi-liquid
temperature regions, reproducing very well also the crossover
between them.

Although the spin-1/2 Anderson model can be solved by using the
exact NRG, we believe that the UNCA is more suitable than NRG for
extension to realistic systems, e.g. for taking into account the
effects of the electronic structure of the dots, and thus we point
out that the UNCA method can become an important tool to interpret
the experiments. Possible applications are, for example, in the
explanation of the Kondo effect in quantum dots for integer
spin\cite{sasaki} and the Kondo effect in carbon
nanotubes.\cite{nygard}

\begin{acknowledgments}
We would like to thank S. De Franceschi for helpful discussions
and for a critical reading of the manuscript, and J. Schmalian for
useful correspondence concerning his UNCA code. The numerical
calculations were partly performed at the Computing Center of
Collegio Borromeo in Pavia, whose support is gratefully
acknowledged.
\end{acknowledgments}


\end{multicols}

\begin{references}
\bibitem{nota:gamma}
The definition of $\Gamma$ given in the present paper is the same
adopted in Ref. \onlinecite{glazman}. This definition differs from
the one given in Refs. \onlinecite{meir91,meir92,meir93,meir94} by
a factor of 2.
\bibitem{nota}
Expressing the results as function of $-\varepsilon_0$, with
$\varepsilon_F$ set equal to zero, is equivalent to plotting the
same curves as function of chemical potential at fixed
$\varepsilon_0$ (see e.g. Fig. 6 of Ref. \onlinecite{meir94}).
\end{references}
\end{document}